\documentclass[11pt]{article}

\begin{document}

\noindent{\large\bf Permutation Generation: Two New Permutation
Algorithms}

\bigskip
\noindent{JIE GAO and DIANJUN WANG }

\noindent{\it\small Tsinghua University, Beijing, China}

\bigskip
\noindent {\small{\bf Abstract.} Two completely new algorithms for
generating permutations, shift-cursor algorithm and level
algorithm, and their efficient implementations are presented in
this paper. One implementation of the shift cursor algorithm gives
an optimal solution of the permutation generation problem, and one
implementation of the level algorithm can be used to generate
random permutations.

}

\section*{1. Introduction}

Permutation generation has a history which dates back to the
1650s. After the electronic computer was invented dozens of
algorithms have been published which generates all the
permutations of $S_n$  by a computer. In 1977 Robert Sedgewick [3]
wrote a survey of this problem and compared nearly all permutation
generation methods at that time. As a basic problem of computer
science, it is still an interesting problem to be studied.

Sedgewick [3] concluded in his survey that his implementation of
Heap's method was the fastest permutation generation algorithm for
most computers, and it can be coded that most permutations are
generated using only two store instructions. This is a lower bound
of permutation algorithms. Of course it can not be exceeded but
algorithms which run as fast as it may be written.

Two completely new algorithms for generating permutations, shift-
cursor algorithm and level algorithm, are presented in this paper
and we give some implementations of them. Like Sedgewick's
implementation of Heap's method, one of the implementations of our
first algorithm comes infinitely close to the theoretically
optimal solution of the problem, that is, $2n!$ store
instructions. Also one implementation of the other algorithm can
be used to generate random permutations.

\section*{2. The Algorithms}

Our two algorithms are strongly related. Specifically, each number
in a "level" permutation gives the level of the shift cursor in
the corresponding permutation.

First, we discuss the shift cursor algorithm which generates the
permutations in $S_n$ beginning with the identity permutation
$123\cdots n$. The algorithm uses the first number 1 in 123...n to
partition the n! permutation in n blocks of $(n-1)!$ permutations
as follows. The algorithm first generates $(n-1)!$ permutations
with cursor 1 in $1^{\mbox{\small\rm st}}$ position, the next
$(n-1)!$ permutations with cursor 1 in the $2^{\mbox{\small\rm
nd}}$ position, and so on, until the cursor 1 is in the
$n^{\mbox{\rm\small   th}}$ position. In other words, the cursor
partitions the $n!$ permutations into $n$ blocks, and in each
block the cursor is in a fixed position. We define the level of
the cursor to be the number of blocks determine by the cursor.
Thus the level of cursor 1 is $n$.

The algorithm, while  generating the permutations with the cursor
1 in a fixed position, uses a new cursor to partition each block
of $(n-1)!$ permutations into $n-1$ {\it smaller blocks} and in
each {\it smaller block} the new cursor is in a fixed position. To
find the cursor of a {\it smaller block}, we first ignore the
cursors of the larger blocks from the permutations, and then use
the first number of the first permutation in the block as the
cursor of the {\it smaller block}. For example,  all the level
$n-1$ cursors are the first numbers other than cursor 1 in the
first permutation of each smaller block. This method can be used
until the new blocks contain only one permutation and then all
$n!$ permutations are generated.

     Consider, for example, the case n = 4 whose permutation sequence appears
in Figure 1.  The sequence begins with the permutation 1234.  The
algorithm first uses 1 as the cursor of all $4!$ permutations, and
we get the following 4 blocks (where * stands for an unknown
number):
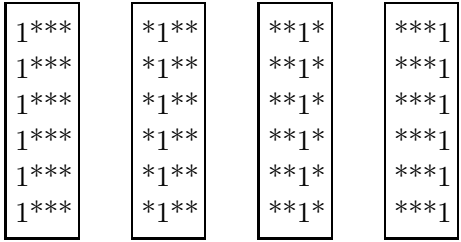
\begin{figure}[htb]
\setlength{\unitlength}{1.2mm}
\begin{picture}(52,28)(0,0)
\put(0,0){\framebox(8,26){\multiput(1,0)(0,4){6}{1***}}}
\put(14,0){\framebox(8,26){\multiput(1,0)(0,4){6}{*1**}}}
\put(28,0){\framebox(8,26){\multiput(1,0)(0,4){6}{**1*}}}
\put(42,0){\framebox(8,26){\multiput(1,0)(0,4){6}{***1}}}
\end{picture}
\end{figure}

\noindent A new cursor is then used in each block to partition the
3! = 6 permutations into 3 smaller blocks. For example, in the
first larger block the first permutation is $1234$, so we use 2 as
the cursor of the first block to get the following 3 smaller
blocks:
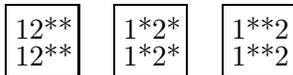
\begin{figure}[htb]
\setlength{\unitlength}{1.2mm}
\begin{picture}(28,10)
\put(0,0){\framebox(8,8){\multiput(1,0)(0,3){2}{12**}}}
\put(12,0){\framebox(8,8){\multiput(1,0)(0,3){2}{1*2*}}}
\put(24,0){\framebox(8,8){\multiput(1,0)(0,3){2}{1**2}}}
\end{picture}
\end{figure}

\noindent  In the first block above, where the first permutation
is 1234, the first number, other than 1 and 2, is 3, so we use 3
as the cursor of the block. Then we get 2 smaller blocks: 123* and
12*3. Both of the blocks contains only one permutation and we put
the last number 4 in the permutations to get 1234 and 1243. We
continue in this way to obtain the other permutation in the first
block ending with the permutation 1432.

In the second block, the first permutation will be 4132.  This
comes from the previous permutation 1432 (the last permutation in
the first block) by interchanging the cursor 1 with the next
number in the permutation. Since 4132 is the first permutation in
the second block, we use the first number 4 as the cursor of the
second block to obtain the following 3 smaller blocks:
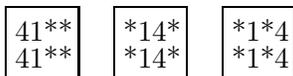
\begin{figure}[htb]
\setlength{\unitlength}{1.2mm}
\begin{picture}(28,10)
\put(0,0){\framebox(8,8){\multiput(1,0)(0,3){2}{41**}}}
\put(12,0){\framebox(8,8){\multiput(1,0)(0,3){2}{*14*}}}
\put(24,0){\framebox(8,8){\multiput(1,0)(0,3){2}{*1*4}}}
\end{picture}
\end{figure}

Continuing as above, we get all the permutations in the second
block. We then get all $n! = 4! = 24$ permutations which appear in
Figure 1.

Observe that, in each permutation, each number in the permutation
is a cursor of a certain block, and in a given permutation the
levels of the cursors are different from each other. Thus the
cursor levels form a permutation. Our level sequence is obtained
from the sequence of cursor level permutations by interchanging 1
and $n$, 2 and $n-1$, and so on.

Figure 1 shows, for $n = 4$, the shift cursor sequence, the levels
of the cursors, and the level sequence.  We also include the
well-known Johnson-Trotter sequence [2] [4] for reference.
($n=4$). See Figure 1.

\begin{center}
\begin{tabular}[htb]{|l|l|l|l|}\hline
shift cursor          &levels\ \ \  \ &level\ \ \ \ &Johnson-Trotter\\
sequence  \ \ & of the cursors & sequence\ \ \ \ \ \ \ &sequence
\\ \hline1234 & 4321 & 1234 &1234\\
\hline 1243&4312&1243&1243\\ \hline 1423&4231&1324&1423\\
\hline 1324&4132&1423&4123\\ \hline 1342&4213&1342&4132\\
\hline
1432&4123&1432&1432\\ \hline & & &\\  \hline 4132&3421&2134&1342\\
\hline 4123&3412&2143&1324\\ \hline 2143&2431&3124&3124\\
\hline 3142&1432&4123&3142\\ \hline 3124&2413&3142&3412\\
\hline
2134&1423&4132&4312\\ \hline & & & \\ \hline 2314&3241&2314&4321\\
\hline 2413&3142&2413&3421\\ \hline 4213&2341&3214&3241\\
\hline 3214&1342&4213&3214\\ \hline 3412&2143&3412&2314\\
\hline
4312&1243&4312&2341\\ \hline & & &\\  \hline 4321&3214&2341&2431\\
\hline 4231&3124&2431&4231\\ \hline 2431&2314&3241&4213\\
\hline 3421&1324&4231&2413\\ \hline 3241&2134&3421&2143\\
\hline 2341&1234&4321&2134\\ \hline & & &\\ \hline
\end{tabular}
\end{center}
\centerline{Figure 1.}

\section*{3. Implementation of the Level Algorithm - Also can be
Used to Generate Random Permutations}

If N is so large that we could never hope to generate all
permutations of N elements, it is of interest to study methods for
generating "random" permutations of N elements. The following
algorithm implements level algorithm by using a number between 1
to $N!$ to generate a permutation of level algorithm. Since we can
generate a "random" number between 1 to $N!$ first and use the
following algorithm to generate a permutation of level algorithm,
the following algorithm can also be used to generate "random"
permutations.

Suppose the function divide(int $a$, int $b$) executes $a/b$ and
put the quotient to $b$, and the remainder to $a$ and $0!=1$. We
can write implementation of level sequence using the following
algorithm:

 {\renewcommand\baselinestretch{1}
 {\tt
 for i=1 to n! step 1 do

 begin

\hspace{.5cm}for j=1 to n step 1 do

 \hspace{.5cm}begin

 \hspace{1cm}int m=(n-j)!

 \hspace{1cm}divide(i, m)

 \hspace{1cm}if i$\not=$0 then

\hspace{1cm}put j to the (m+1)$^{\mbox {\small\tt th}}$ empty
position

\hspace{1cm}else

\hspace{1cm}put j to the m$^{\mbox{\small\tt th}}$ empty position

\hspace{.5cm}end

\hspace{.5cm}output one permutation

end} }

For example, suppose $n=4$, to generate the 15$^{\mbox{\rm\small
th}}$ permutation we first calculate $15/3!=15/6=2\cdots3$, the
quotient is 2 the remainder is 3, then we put number 1 in the
3$^{\mbox{\rm\small   rd}}$ position, we get **1*. Secondly we
calculate $3/2!=3/2=1\cdots1$, the quotient is 1 the remainder is
1, then we put number 2 in the 2$^{\mbox{\rm\small   nd}}$ empty
position, we get *21*. Third we calculate $1/1!=1\cdots 0$, the
quotient is 1 the remainder is 0, then the number 3 is in the
1$^{\mbox{\rm\small   st}}$ empty position, we get 321*. Finally
we calculate 0/0!=0/1=0...0, the quotient is 0 the remainder is 0,
then the number 4 is in the 1$^{\mbox{\rm\small   st}}$ empty
position, we get the permutation 3214 at last.

\section*{4. Implementation of Shift Cursor Algorithm - an Optimal
Solution of Permutation Generation Problem}

 It is can be noticed that shift cursor sequence can generate next permutation by
transposition. And the sequence has the following properties:

{\bf Property 1. }{\it A level $j$ cursor remains the same
position in continuous $(j-1)!$ permutations.}

{\bf Property 2. }{\it If ignore the cursors which level $>j$ from
a sequence of $n!$ permutations then the sequence turns to $n!/j!$
blocks with length $j$, and the rule of change of each blocks is
the same.}

We call Property 1 as the local stability of the sequence, and
call Property 2 as the rule stability of the sequence.

Property 1 ensures that when generating permutations using shift
cursor algorithm, cursors with higher levels can be placed in
fixed positions for a certain part of the sequence.

Property 2 ensures that for the certain part of the sequence, if
higher level cursors have already been placed in their positions,
a {\it predetermined} rule can be used to place other cursors in
their right positions.

Clearly [1], there are two types of operations needed for
generating a new permutation from the previous permutation:
\begin{enumerate}
\item Operations needed for actually {\it executing changes} from
the previous permutation. \item Operations needed for {\it
reaching decisions} about which type 1 operations are required.
\end{enumerate}

In theory, any permutation generating algorithm should contain the
two types of operations and only the two types of operations, and
a theoretically optimal algorithm for the problem should has the
following factors:
\begin{enumerate}
\item[1.]All type 1 operations should be change only two numbers of
the permutation.
\item[2.]The number of type 2 operations should be limited to reach
zero, or much smaller than type 1 operations.
\end{enumerate}
By using the two properties of shift cursor sequence, algorithms
which have the above two factors can be written.

Suppose we have already written an algorithm for generating
permutations of shift cursor algorithm. For a certain part of the
sequence, such as a sequence of continuous $j!$ permutations, we
can place the cursors which level $>j$ in their places first, and
these cursors remain the fixed positions in the $j!$ permutations
(property 1). Then by using a given rule, we can place other
cursors directly (property 2), without any type 2 operations.

For example, suppose we are using the algorithm to generate 5!
permutations $(n=5)$, and suppose $j=3$. Starting from 12345, the
level of number 1 is 5, the level of number 2 is 4. Then we fix
number 1 in the 1$^{\mbox{\rm\small   st}}$ position, fix number 2
in the 2$^{\mbox{\rm\small   nd}}$ position. Afterwards, by using
a given rule, we can place the numbers 3, 4, 5 directly in their
positions to generate permutations. The rule to place the three
numbers is the same as the rule to generate 3! permutations
$(n=3)$ using shift cursor algorithm. When we using the algorithm
to generate $3!$ permutations ($n=3$), the sequence is as the
following: 123, 132, 312, 213, 231, 321. In this case we name 3
the No.1 element, name 4 the No.2 element, name 5 the No.3
element. In the above sequence, No.2 element and No.3 element be
changed firstly, so we change 4, 5 first, then the next
permutation to 12345 is 12354. And No.1 element and No.3 element
changed afterwards, so we change 3, 5 secondly, then the next
permutation to 12354 is 12534, and so on. And finally we get the
6$^{\mbox{\rm\small   th}}$ permutation 12543. Some type 2
operations is needed now to determine which cursor's level $>j$
and where should we place these cursors. After that, we can
generate next $3!-1=5$ permutations by directly change two
numbers. This algorithm can be described as following:

Algorithm 1: {\tt

1.Starting from 1, 2, 3, ... , n.

2.If the level order of the cursors is 1, 2, 3, ... , n, stop.

3.Determine cursors which level>j and their positions.

4.Output.

5.Generate and output next j!-1 permutations directly.

6.Go to step 2. }

When generating permutations using algorithm 1, there are
$(j!-1)/j!$ permutations are generated by directly change two
numbers without any type 2 operations. And other $1/j!$
permutations are generated using additional type 2 operations.

Suppose when using algorithm 1, the average execution time of type
1 operation is t and the average execution time of type 2
operation is T, then the overall execution time of algorithm 1
(T1) is $n! * t + n!/j! * T$. Since every type 1 operation of
algorithm 1 is change two numbers, the overall execution time of
theoretically optimal algorithm of permutation generation (T0) is
$n! * t$, and T1/T0$= 1 + T / (j! * t)>1$. So with $j$ grows
larger T1/T0 could infinitely close to 1, which means algorithm 1
could infinitely close to the theoretically optimal solution of
the problem.

Robert Sedgewick's implementation of Heap's method is very similar
to algorithm 1, the only difference between them is the
termination condition (step 2) of the algorithms, and they both
have the property 1 and property 2. And it is the two properties
make the two algorithms both can infinitely close to the
theoretically optimal solution of the problem.

\section*{5. Summary and Conclusions}

Although there are few applications need to generate all n!
permutations, permutation generation is still an interesting
problem to be studied. Further studies are needed on this problem,
and these studies may change our understanding of permutations in
the future.

Comparing with permutation generation, random permutation
generation has more practical meanings. The implementation of
level algorithm gives a new method to generate random permutation.
And it can be used to generate the permutation which in a given
position of the level sequence.

Because of the two important properties, the shift cursor
algorithm has an efficient implementation, and in theory both the
implementation and Sedgewick's implementation of Heap's method
could infinitely close to the theoretically optimal solution of
the problem, which is $2n!$ store instructions. Comparing with
Heap's algorithm and another famous algorithm -- Johnson-Trotter
algorithm, the shift cursor algorithm is easier to understand than
Heap's algorithm and has a more efficient implementation than
Johnson-Trotter algorithm. And it's more suitable to be written in
textbooks.

\bigskip
\textbf{Acknowledgement.} Appreciation must be expressed to
Professor Seymour Lipschutz for his valuable comments and
suggestions for improving the readability of the manuscript.

\bigskip
\noindent {\bf Jie Gao} ({\tt gj02@mails.tsinghua.edu.cn}) is a
MSE student of School of Software, Tsinghua University, Beijing,
China.

\noindent {\bf Dianjun Wang} ({\tt djwang@math.tsinghua.edu.cn})
is an associate professor of Department of Mathematical Sciences,
Tsinghua University, Beijing, China.

\end{document}